\documentclass[10pt]{article}
\usepackage{amsfonts}
\usepackage{graphics,epsfig}

\newtheorem{definition}{Definition}
\newtheorem{proposition}{Proposition}

\usepackage[psamsfonts]{amssymb}

\newcommand{\be}{\begin{equation}}
\newcommand{\ee}{\end{equation}}
\newcommand{\bea}{\begin{eqnarray}}
\newcommand{\eea}{\end{eqnarray}}
\newcommand{\ba}{\begin{array}}
\newcommand{\ea}{\end{array}}

\newcommand{\bp}{\begin{proposition}}
\newcommand{\ep}{\end{proposition}}
\newcommand{\bd}{\begin{definition}}
\newcommand{\ed}{\end{definition}}

\begin{document}

\title{Entanglement as Internal Constraint}
\author{Diederik Aerts, Ellie D'Hondt and Bart D'Hooghe}

\date{}
\maketitle

\centerline{Center Leo Apostel, Brussels Free University}
\centerline{Krijgskundestraat 33, 1160 Brussels, Belgium}
\centerline{diraerts,eldhond,bdhooghe@vub.ac.be}

\begin{abstract}
Our investigation aims to study the specific role played by entanglement in
the quantum computation process, by elaborating an entangled spin model
developed within the `hidden measurement approach' to quantum mechanics. We
show that an arbitrary tensor product state for the entity consisting of two
entangled qubits can be described in a complete way by a specific internal
constraint between the ray and density states of the two qubits. For the
individual qubits we use a sphere model representation, which is a
generalization of the Bloch or Pauli representation, where also the collapse
and noncollapse measurements are represented. We identify a parameter $r\in
[0,1]$, arising from the Schmidt diagonal decomposition, that is a measure
of the amount of entanglement, such that for $r={0}$ the system is in the
singlet state with `maximal' entanglement, and for $r=1$ the system is in a
pure product state.
\end{abstract}


\section{Introduction}

In quantum computation the concepts of quantum superposition states and
quantum entanglement are crucial. We want to study quantum entanglement in
the most simple case, namely a system consisting of two entangled spin $%
\frac 12.$ The quantum entity consisting of two entangled spin $\frac 12$ is
described in the tensor product of the two dimensional complex Hilbert
spaces that describe the single spins. Let us refer to the first spin as the
`left spin' and to the second spin as the `right spin'. It is well known
that for the two spins being in the singlet state, the typical EPR
correlations are encountered, meaning that if the left spin collapses in a
certain direction under the influence of a measurement, then the right spin
collapses in the opposite direction. 

Our aim is to study in detail the entanglement for an arbitrary tensor
product state that is not necessarily the singlet state, by making use of
the sphere model representation for the spin of a spin ${\frac 12}$ particle
that was developed in Brussels within the `hidden measurement approach' to
quantum mechanics \cite{Aerts1986,Aerts1987,Aerts1991,aerts94a,aerts94b,aerts1997,aerts1997b,dhooghe2000}. We do this by introducing `constraint functions' that
describe the behavior of the state of one of the spins if measurements are
executed on the other spin.

We will consider two types of measurements: (1) noncollapse measurements, of
which the action on a mixture of states is described by Luder's formula, and
(2) collapse measurements, of which the action is described by Von Neumann's
formula. We will show that (1) an arbitrary noncollapse measurement on one
of the two spins does not provoke any change in the partial trace density
matrix of the other spin, i.e., the spins behave as separated entities for
noncollapse measurements; (2) an arbitrary collapse measurement on one spin
provokes a rotation and a stretching on the other spin, which can be
described in detail by means of the sphere model.

\section{The Sphere Model}

In the sphere model representation a ray state of the spin is represented by
a point of a sphere with radius 1 in the three dimensional real space ${%
\mathbb R}^3$, such that the direction of the point towards the origin of
the sphere coincides with the direction of the spin in three dimensional
space as measured for example by a Stern-Gerlach apparatus.

Let us denote the point $(r\sin \theta \cos \phi ,r\sin \theta \sin \phi
,r\cos \theta )$ of ${\mathbb R}^3$ by the vector $u(r,\theta ,\phi )$. The
ray state 
\begin{equation}
|\theta \phi \rangle =(\cos {\frac \theta 2}e^{-i{\frac \phi 2}},\sin {\frac
\theta 2}e^{\frac{i\phi }2})
\end{equation}
vector of ${\mathbb C}^2$, is then represented by 
\begin{equation}
u(1,\theta ,\phi )=(\sin \theta \cos \phi ,\sin \theta \sin \phi ,\cos
\theta )
\end{equation}
vector of ${\mathbb R}^3$, and point of the sphere with radius 1 and center
in $(0,0,0)$. We remark that this part of our sphere model is nothing else
but the well known Poincar\'{e} representation of ${\mathbb C}^2$. A density
state of the spin is represented by an interior point of the sphere, which
is a convex linear combination of points of the surface of the sphere, in
such a way that the weights of the convex combination coincide with the
weights of the statistical mixture that corresponds with the density state.
It is not difficult to calculate the density state $D(r,\theta ,\phi )$ that
corresponds with an arbitrary interior point, $u(r,\theta ,\phi )$, $r\in
[0,1],\theta \in [0,\pi ],\phi \in [0,2\pi ]$, of the sphere.

To do this we remark that also a ray state has a density representation,
where in this case the density matrix is the orthogonal projection on the
ray. This means that the density matrix representing the ray state $|\theta
\phi \rangle $ is given by: 
\begin{eqnarray}
D(1,\theta ,\phi )=|\theta \phi \rangle \langle \theta \phi | &=&\left( 
\begin{array}{cc}
\cos ^2{\frac \theta 2} & \cos {\frac \theta 2}\sin {\frac \theta 2}%
e^{-i\phi } \\ 
\cos {\frac \theta 2}\sin {\frac \theta 2}e^{i\phi } & \sin ^2{\frac \theta 2%
}
\end{array}
\right)  \\
&=&{\frac 12}\left( 
\begin{array}{cc}
1+\cos \theta  & \sin \theta e^{-i\phi } \\ 
\sin \theta e^{i\phi } & 1-\cos \theta 
\end{array}
\right) 
\end{eqnarray}
The ray state orthogonal to $|\theta \phi \rangle $ is $|\pi -\theta ,\phi
+\pi \rangle $, and this state is represented by the point $-u(1,\theta
,\phi )$ of the sphere, corresponding to the opposite spin direction. We
have: 
\begin{eqnarray}
D(1,\pi -\theta ,\phi +\pi ) &=&|\pi -\theta ,\pi +\phi \rangle \langle \pi
-\theta ,\pi +\phi | \\
&=&{\frac 12}\left( 
\begin{array}{cc}
1-\cos \theta  & -\sin \theta e^{-i\phi } \\ 
-\sin \theta e^{i\phi } & 1+\cos \theta 
\end{array}
\right) 
\end{eqnarray}
To find the general representation for $u(r,\theta ,\phi )$ we remark that
the center of the sphere, hence the point $u(0,\theta ,\phi )=(0,0,0)$, can
be written as the convex combination ${\frac 12}u(1,\theta ,\phi )+{\frac 12}%
u(1,\pi -\theta ,\pi +\phi )$. This means that the density matrix that
represent the center of the sphere is given by: 
\begin{equation}
D(0,\theta ,\phi )={\frac 12}D(1,\theta ,\phi )+{\frac 12}D(1,\pi -\theta
,\pi +\phi )=\left( 
\begin{array}{cc}
\frac 12 & 0 \\ 
0 & \frac 12
\end{array}
\right) 
\end{equation}
We further have that: 
\begin{equation}
u(r,\theta ,\phi )=ru(1,\theta ,\phi )+(1-r)u(0,\theta ,\phi )
\end{equation}
and hence: 
\begin{eqnarray}
D(r,\theta ,\phi ) &=&rD(1,\theta ,\phi )+(1-r)D(0,\theta ,\phi ) \\
&=&{\frac 12}\left( 
\begin{array}{cc}
1+r\cos \theta  & r\sin \theta e^{-i\phi } \\ 
r\sin \theta e^{i\phi } & 1-r\cos \theta 
\end{array}
\right) 
\end{eqnarray}
which gives us the representation of a general density state $D(r,\theta
,\phi )$ by means of the interior point $u(r,\theta ,\phi )$ of the sphere (see Figure 1).

\begin{figure}[htpb]
\centering
\includegraphics{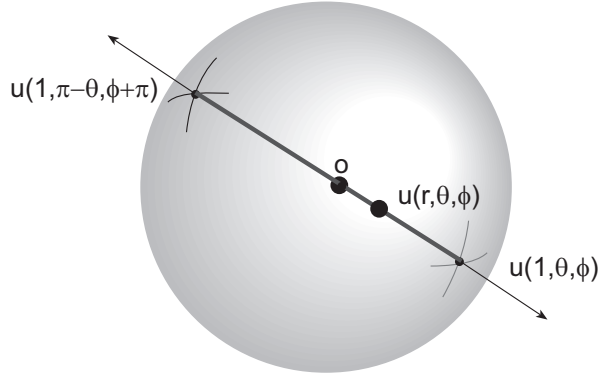}
\caption{Representation of a general density state 
$D(r,\theta,\phi )$ by means of the interior point $u(r,\theta ,\phi )$ of the sphere.}
\end{figure}

We can see that the part of the sphere model that we developed in Brussels
that relates to the representation of the density and ray states of the
spin, is the Bloch representation. Additionally to this state part of the
representation, we however also developed a representation for the
measurements in our sphere model. Before we explain the way in which
measurements are represented, let us identify some of the special points of
this sphere representation. We have seen already that the center of the
sphere, hence the point $u(0,\theta ,\phi )$ represents that density state 
\begin{equation}
D(0,\theta ,\phi )=\left( 
\begin{array}{cc}
\frac 12 & 0 \\ 
0 & \frac 12
\end{array}
\right) 
\end{equation}
The North pole of the sphere, hence the point $u(1,0,\phi )$, represents the
state 
\begin{equation}
D(1,0,\phi )=\left( 
\begin{array}{cc}
1 & 0 \\ 
0 & 0
\end{array}
\right) 
\end{equation}
which is the orthogonal projector on the first canonical base vector $(1,0)$
of ${\mathbb C}^2$, while the South of the sphere, hence the point $u(1,\pi
,\phi )$, represents the state 
\begin{equation}
D(1,\pi ,\phi )=\left( 
\begin{array}{cc}
0 & 0 \\ 
0 & 1
\end{array}
\right) 
\end{equation}
which is the orthogonal projector on the second canonical base vector $(0,1)$
of ${\mathbb C}^2$. An arbitrary point of the straight line connecting the
North pole with the South pole of the sphere, hence $u(r,0,\phi )$ or $%
u(r,\pi ,\phi )$, represents the density states 
\begin{eqnarray}
D(r,0,\phi ) &=&{\frac 12}\left( 
\begin{array}{cc}
1+r & 0 \\ 
0 & 1-r
\end{array}
\right)  \\
D(r,\pi ,\phi ) &=&{\frac 12}\left( 
\begin{array}{cc}
1-r & 0 \\ 
0 & 1+r
\end{array}
\right) 
\end{eqnarray}

Without loss of generality we can demonstrate the effect of a measurement by
considering states that are on the straight line connecting the North and
the South pole of the sphere. So, suppose that the spin is in density state $%
D(r,0,0)$, and that a measurement of the spin is executed with a Stern
Gerlach apparatus in the direction $u(1,\theta ,\phi )$. Quantum mechanics
prescribes the way, by means of Luder's formula, in which we calculate the
density state of the spin after this measurement. 
\begin{equation}
D=P(\theta ,\phi )D(r,0,0)P(\theta ,\phi )+(1-P(\theta ,\phi
))D(r,0,0)(1-P(\theta ,\phi ))
\end{equation}
where $P(\theta ,\phi )$ is the projector on the ray state $|\theta \phi
\rangle $. We know that $P(\theta ,\phi )=D(1,\theta ,\phi )$. If we make
this calculation we find 
\begin{equation}
D={\frac 12}\left( 
\begin{array}{cc}
1+r\cos ^2\theta  & r\sin \theta \cos \theta e^{-i\phi } \\ 
r\sin \theta \cos \theta e^{i\phi } & 1-r\cos ^2\theta 
\end{array}
\right) 
\end{equation}
Let us see which point of the sphere corresponds with this density state. To
do this, let us first suppose that $\theta \in [0,{\frac \pi 2}]$. In this
case we can introduce 
\begin{equation}
r^{\prime }=r\cos \theta 
\end{equation}
and we have 
\begin{equation}
D={\frac 12}\left( 
\begin{array}{cc}
1+r^{\prime }\cos \theta  & r^{\prime }\sin \theta e^{-i\phi } \\ 
r^{\prime }\sin \theta e^{i\phi } & 1-r^{\prime }\cos \theta 
\end{array}
\right) =D(r^{\prime },\theta ,\phi )
\end{equation}
This means that for $\theta \in [0,{\frac \pi 2}]$ we have that $D(r,0,0)$
transform into $D(r\cos \theta ,\theta ,\phi )$, if a measurement with a
Stern Gerlach in direction $(\theta ,\phi )$ is executed. Consider now the
case where $\theta \in [{\frac \pi 2},\pi ]$. We can put then 
\begin{eqnarray}
r^{\prime } &=&r\cos (\pi -\theta ) \\
\theta ^{\prime } &=&\pi -\theta  \\
\phi ^{\prime } &=&\phi +\pi 
\end{eqnarray}
and find 
\begin{equation}
D={\frac 12}\left( 
\begin{array}{cc}
1+r^{\prime }\cos \theta ^{\prime } & r^{\prime }\sin \theta ^{\prime
}e^{-i\phi ^{\prime }} \\ 
r^{\prime }\sin \theta ^{\prime }e^{i\phi ^{\prime }} & 1-r^{\prime }\cos
\theta ^{\prime }
\end{array}
\right) =D(r^{\prime },\theta ^{\prime },\phi ^{\prime })
\end{equation}
which means that for $\theta \in [{\frac \pi 2},\pi ]$ the density state $%
D(r,0,0)$ transforms into $D(r^{\prime },\theta ^{\prime },\phi ^{\prime })$%
, if a measurement with Stern Gerlach in direction $(\theta ,\phi )$ is
executed. If we consider the sphere we can see easily that in both cases the
point $u(r,0,0)$ is transformed into the point 
\begin{equation}
(u(r,0,0)\cdot u(1,\theta ,\phi ))u(1,\theta ,\phi )
\end{equation}
where $u(r,0,0)\cdot u(1,\theta ,\phi )$ is the scalar product in ${\mathbb R%
}^3$ of the vectors $u(r,0,0)$ and $u(1,\theta ,\phi )$. This means that we
have identified a very simple mechanics to describe the quantum measurement
effect in our sphere model. The effect is just an ordinary orthogonal
projection on the direction of the Stern Gerlach apparatus of the point that
represents that ray or density state of the spin in the sphere model (see Figure 2).

\begin{figure}[htpb]
\centering
\includegraphics{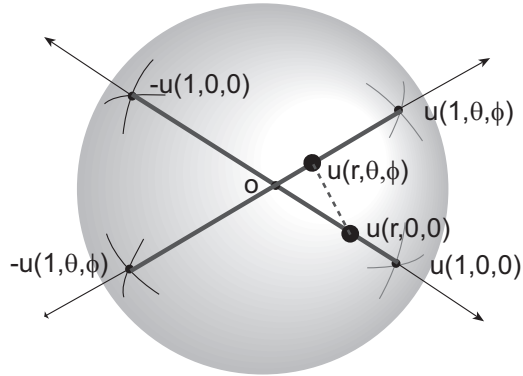}
\caption{Effect of the measurement on a single spin $\frac 12$.}
\end{figure}

Let us formulate the general case. Suppose that we have a spin state
represented by the point $u(s,\alpha ,\beta )$ and we perform a measurement
with a Stern Gerlach apparatus in direction $(\theta ,\phi )$. We denote the
orthogonal projection on the straight line with direction $(\theta ,\phi )$
in ${\mathbb R}^3$ by $E(\theta ,\phi )$. Then the new state after a quantum
mechanical measurement with Stern Gerlach in direction $(\theta ,\phi )$,
when the state of the spin before the measurement is represented in the
sphere model by the point $u(s,\alpha ,\beta )$, is given by 
\begin{equation}
E(\theta ,\phi )u(s,\alpha ,\beta )
\end{equation}
and we have 
\begin{eqnarray}
E(\theta ,\phi )u(s,\alpha ,\beta ) &=&u(s\cos \theta ,\theta ,\phi )\ 
\mathrm{if}\ |\alpha -\theta |\in [0,{\frac \pi 2}] \\
E(\theta ,\phi )u(s,\alpha ,\beta ) &=&u(s\cos (\pi -\theta ),\pi -\theta
,\phi +\pi )\ \mathrm{if}\ |\alpha -\theta |\in [{\frac \pi 2},\pi ]
\end{eqnarray}

It is possible to give a nice geometrical presentation of how the spin state
changes under the influence of measurements in different directions (see Figure 3).

\begin{figure}[htpb]
\centering
\includegraphics{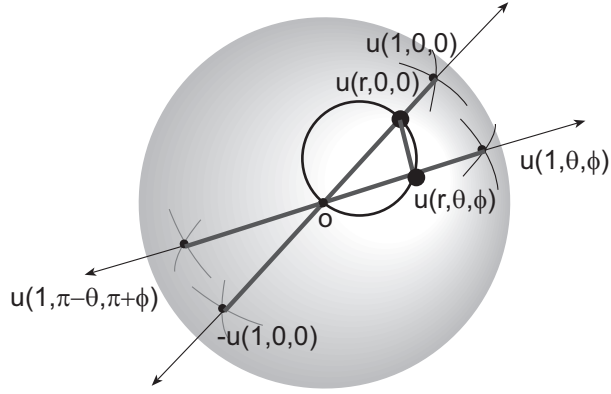}
\caption{A geometrical presentation of how the spin state
changes under the influence of measurements in different directions.}
\end{figure}

Consider a little sphere inside the big sphere of the model, such that the
North pole of the little sphere is in the point $u(s,\alpha ,\beta )$, the
point that represents the spin state, and the South pole is in the center of
the big sphere. Consider now a straight line with direction $(\theta ,\phi )$
through the center of the big sphere, representing the direction of the
Stern Gerlach apparatus. The point where this line cuts the little sphere is
the point where the spin state will be transformed to under influence of the
measurement. This also means that the points of the little sphere are the
points that represent the states where under arbitrary angles for the
measurement the spin state can be transformed to.


\section{Entangled Spins}

The entity consisting of two entangled spin ${\frac{1 }{2}}$ is described by
means of the tensorproduct ${\mathbb C}^2_1 \otimes {\mathbb C}^2_2$, where $%
{\mathbb C}_1$ and ${\mathbb C}_2$ are two copies of ${\mathbb C}$, that we
label with indices $1$ and $2$ with the sole purpose of identifying them.
This means that the ray states of this entangled spin ${\frac{1 }{2}}$
entity are described by the rays of ${\mathbb C}^2_1 \otimes {\mathbb C}^2_2$
and the density states by the density matrices of ${\mathbb C}^2_1 \otimes {%
\mathbb C}^2_2$.

\subsection{The Constraint Functions}

Suppose that we consider an arbitrary unit vector $\psi \in {\mathbb C}%
_1^2\otimes {\mathbb C}_2^2$. Then it is always possible to write $\psi $ as
the sum of product vectors 
\begin{equation}
\psi =\sum_{ij}\lambda _{ij}e_1^i\otimes e_2^j
\end{equation}
where $\lambda _{ij}\in {\mathbb C}$, and $\{e_1^i\}$ and $\{e_2^j\}$ are a
bases respectively of ${\mathbb C}_1^2$ and ${\mathbb C}_2^2$.

Let us consider a measurement on the first spin. This measurement provokes
the first spin to collapse with a certain probability into a spin state
described by a unit vector $x_1\in {\mathbb C}_1^2$. The state $\psi $ of
the entangled spins is transformed in the state 
\begin{equation}
(P_{x_1}\otimes I)(\psi )
\end{equation}
where $P_{x_1}$ is the orthogonal projector of ${\mathbb C}_1^2$ on $x_1$,
and $I$ is the unit operator of ${\mathbb C}_2^2$. The result is that the
entangled spins end up in a product state that is the following: 
\begin{eqnarray}
(P_{x_1}\otimes I)(\psi ) &=&\sum_{ij}\lambda _{ij}(P_{x_1}\otimes
I)(e_1^i\otimes e_2^j) \\
&=&\sum_{ij}\lambda _{ij}\langle x_1,e_1^i\rangle x_1\otimes e_2^j \\
&=&x_1\otimes \sum_{ij}\lambda _{ij}\langle x_1,e_1^i\rangle e_2^j
\end{eqnarray}
This means that as a consequence of the spin measurement on the first spin,
making its state collapse in the state $x_1$, the spin state of the second
spin collapses to the state 
\begin{equation}
\sum_{ij}\lambda _{ij}\langle x_1,e_1^i\rangle e_2^j
\end{equation}
In an analogous way we can show that if a measurement is performed on the
second spin that makes its state collapse to the state described by the unit
vector $x_2\in {\mathbb C}_2^2$, the state of the first spin collapses to
the state described by the vector 
\begin{equation}
\sum_{ij}\lambda _{ij}\langle x_2,e_2^j\rangle e_1^i
\end{equation}

\begin{definition}[Constraint Functions]
Let us consider the functions $F_{12}(\psi )$ and $F_{21}(\psi )$ defined in
the following way 
\begin{eqnarray}
F_{12}(\psi ) &:&{\mathbb C}_1^2\rightarrow {\mathbb C}_2^2:x_1\mapsto
\sum_{ij}\lambda _{ij}\langle x_1,e_1^i\rangle e_2^j \\
F_{21}(\psi ) &:&{\mathbb C}_2^2\rightarrow {\mathbb C}_1^2:x_2\mapsto
\sum_{ij}\lambda _{ij}\langle x_2,e_2^j\rangle e_1^i
\end{eqnarray}
We call $F_{12}(\psi )$ and $F_{21}(\psi )$ the constraint functions related
to $\psi $.
\end{definition}

These constraint functions map the unit vectors describing the state where
the entangled spin collapses to by a measurement on one of the spins to the
vector describing the state that the other spin collapses to under influence
of the entanglement correlation. A detailed study of the constraint
functions can give us a complete picture of how the entanglement correlation
works as an internal constraint. Before we arrive at this complete picture,
let us proof some properties of the constraint functions that we need to
derive the picture.

\begin{proposition}
The constraint functions are canonically defined
\end{proposition}

Proof: Indeed, consider other bases $\{f_1^k\}$ of ${\mathbb C}_1^2$, and $%
\{f_2^l\}$ of ${\mathbb C}_2^2$, such that 
\begin{equation}
\psi =\sum_{kl}\mu _{kl}f_1^k\otimes f_2^l
\end{equation}
We have 
\begin{eqnarray}
f_1^k &=&\sum_ia_i^ke_1^i \\
f_2^l &=&\sum_jb_j^le_2^j
\end{eqnarray}
and hence 
\begin{equation}
\psi =\sum_{klij}\mu _{kl}a_i^kb_j^le_1^i\otimes e_2^j
\end{equation}
From this follows that 
\begin{equation}
\lambda _{ij}=\sum_{kl}\mu _{kl}a_i^kb_j^l
\end{equation}
Hence we have 
\begin{eqnarray}
\sum_{kl}\mu _{kl}\langle x_1,f_1^k\rangle f_2^l &=&\sum_{klij}\mu
_{kl}a_i^kb_j^l\langle x_1,e_1^i\rangle e_2^j \\
&=&\sum_{ij}\lambda _{ij}\langle x_1,e_1^i\rangle e_2^j \\
&=&F_{12}(\psi )(x_1)
\end{eqnarray}
which proves that the definition of $F_{12}(\psi )$ does not depend on the
chosen bases. In an analogous way we prove that $F_{21}(\psi )$ is
canonically defined. \mbox{} \hfill $\Box $

\begin{proposition}
The constraint functions are conjugate linear
\end{proposition}

Proof: Consider $x_1, y_1 \in {\mathbb C}^2_1$ and $\lambda \in {\mathbb C}$%
. We have 
\begin{eqnarray}
F_{12}(\psi)(x_1 + \lambda y_1) &=& \sum_{ij}\lambda_{ij} \langle x_1 +
\lambda y_1, e_1^i \rangle e_2^j \\
&=& \sum_{ij}\lambda_{ij} (\langle x_1, e_1^i \rangle + \lambda^* \langle
y_1, e_1^i \rangle) e_2^j \\
&=& \sum_{ij} \lambda_{ij} \langle x_1, e_1^i \rangle e_2^j + \lambda^*
\sum_{ij} \lambda_{ij} \langle y_1, e_1^i \rangle e_2^j \\
&=& F_{12}(\psi)(x_1) + \lambda^*F_{12}(\psi)(y_1)
\end{eqnarray}
The conjugate linearity of $F_{21}(\psi)$ is proven in an analogous way %
\mbox{} \hfill $\Box$

\bigskip \noindent
Let us calculate $F_{21}(\psi) \circ F_{12}(\psi)$ and $F_{12}(\psi) \circ
F_{21}(\psi)$.

\begin{proposition}
We have 
\begin{eqnarray}
D_1(\psi ) &=&F_{21}(\psi )\circ F_{12}(\psi ) \\
D_2(\psi ) &=&F_{12}(\psi )\circ F_{21}(\psi )
\end{eqnarray}
where $D_1(\psi )$ is the partial trace density matrix to ${\mathbb C}_1^2$
and $D_2(\psi )$ is the partial trace density matrix to ${\mathbb C}_2^2$.
\end{proposition}

Proof: Let us first calculate $D_1(\psi )$ directly. We have 
\begin{equation}
|\psi \rangle \langle \psi |=\sum_{ijkl}\lambda _{ij}\lambda
_{kl}^{*}|e_1^i\rangle \langle e_1^k|\otimes |e_2^j\rangle \langle e_2^l|
\end{equation}
and hence 
\begin{eqnarray}
D_1(\psi ) &=&\sum_{ijkl}\lambda _{ij}\lambda _{kl}^{*}\langle
e_2^j|e_2^l\rangle |e_1^i\rangle \langle e_1^k| \\
&=&\sum_{ijkl}\lambda _{ij}\lambda _{kl}^{*}\delta ^{jl}|e_1^i\rangle
\langle e_1^k| \\
&=&\sum_{ijk}\lambda _{ij}\lambda _{kj}^{*}|e_1^i\rangle \langle e_1^k|
\end{eqnarray}
This means that 
\begin{equation}
D_1(\psi )(x_1)=\sum_{ijk}\lambda _{ij}\lambda _{kj}^{*}\langle
e_1^k,x_1\rangle e_1^i
\end{equation}
Let us now calculate $F_{21}(\psi )\circ F_{12}(\psi )(x_1)$. We have 
\begin{eqnarray}
F_{21}(\psi )\circ F_{12}(\psi )(x_1) &=&\sum_{kl}\lambda _{kl}\langle
F_{12}(\psi )(x_1),e_2^l\rangle e_1^k \\
&=&\sum_{klij}\lambda _{kl}\lambda _{ij}^{*}\langle e_1^i,x_1\rangle \langle
e_2^j,e_2^l\rangle e_1^k \\
&=&\sum_{klij}\lambda _{kl}\lambda _{ij}^{*}\langle e_1^i,x_1\rangle \delta
^{jl}e_1^k \\
&=&\sum_{kli}\lambda _{kl}\lambda _{il}^{*}\langle e_1^i,x_1\rangle e_1^k
\end{eqnarray}
This proves that 
\begin{equation}
F_{21}(\psi )\circ F_{12}(\psi )=D_1(\psi )
\end{equation}
In an analogous way we prove 
\begin{equation}
F_{12}(\psi )\circ F_{21}(\psi )=D_2(\psi )
\end{equation}
\mbox{} \hfill $\Box $

\begin{proposition}
The constraint functions are related in the following way. For $x_1\in {%
\mathbb C}_1^2$ and $x_2\in {\mathbb C}_2^2$ we have 
\begin{equation}
\langle F_{12}(\psi )(x_1),x_2\rangle =\langle x_1,F_{21}(\psi )(x_2)\rangle
^{*}  \label{eq:constraintrelated}
\end{equation}
\end{proposition}

Proof: We have 
\begin{equation}
\langle F_{12}(\psi)(x_1), x_2 \rangle = \langle \sum_{ij}\lambda_{ij}
\langle x_1, e_1^i \rangle e_2^j, x_2 \rangle =
\sum_{ij}\lambda^*_{ij}\langle x_1, e_1^i \rangle^*\langle e_2^j, x_2 \rangle
\end{equation}
and 
\begin{equation}
\langle x_1, F_{21}(\psi)(x_2) \rangle = \langle x_1, \sum_{ij}\lambda_{ij}
\langle x_2, e_2^j \rangle e_1^i \rangle = \sum_{ij}\lambda_{ij} \langle
x_2, e_2^j \rangle \langle x_1, e_1^i \rangle
\end{equation}
\mbox{} \hfill $\Box $

To derive a complete view of how
the entanglement between the two spins works as an internal constraint, let
us derive the way in which the Schmidt diagonal form is related to the
constraint functions.

\subsection{The Schmidt Diagonal Form}

It is always possible to choose a base in ${\mathbb C}_1^2$ and a base in ${%
\mathbb C}_2^2$ such that $\psi $ becomes very simple. This special form for 
$\psi $ is often called the Schmidt diagonalization form. Let us explain how
this works. Since $D_1(\psi )$ is a density matrix, it is of the form

\begin{equation}
D_1(\psi )={\frac 12}\left( 
\begin{array}{cc}
1+r\cos \theta & r\sin \theta e^{-i\phi } \\ 
r\sin \theta e^{i\phi } & 1-r\cos \theta
\end{array}
\right)
\end{equation}
We choose the base 
\begin{eqnarray}
x_1^1 &=&(\cos {\frac \theta 2}e^{-i{\frac \phi 2}},\sin {\frac \theta 2}e^{%
\frac{i\phi }2}) \\
x_1^2 &=&(-i\sin {\frac \theta 2}e^{-i{\frac \phi 2}},i\cos {\frac \theta 2}%
e^{\frac{i\phi }2})
\end{eqnarray}
then in this new base, we have 
\begin{equation}
D_1(\psi )={\frac 12}\left( 
\begin{array}{cc}
1+r & 0 \\ 
0 & 1-r
\end{array}
\right)
\end{equation}
Define now 
\begin{eqnarray}
x_2^1 &=&{\frac{\sqrt{2}}{\sqrt{1+r}}}F_{12}(\psi )(x_1^1) \\
x_2^2 &=&{\frac{\sqrt{2}}{\sqrt{1-r}}}F_{12}(\psi )(x_1^2)
\end{eqnarray}
We have then: 
\begin{eqnarray}
\Vert x_2^1\Vert ^2 &=&{\frac 2{1+r}}\langle F_{12}(\psi
)(x_1^1),F_{12}(\psi )(x_1^1)\rangle \\
&=&{\frac 2{1+r}}\langle x_1^1,F_{21}(\psi )\circ F_{12}(\psi
)(x_1^1)\rangle ^{*} \\
&=&{\frac 2{1+r}}\langle x_1^1,{\frac{1+r}2}x_1^1\rangle ^{*} \\
&=&1
\end{eqnarray}
and 
\begin{eqnarray}
D_2(\psi )(x_2^1) &=&{\frac{\sqrt{2}}{\sqrt{1+r}}}F_{12}(\psi )\circ
F_{21}(\psi )\circ F_{12}(\psi )(x_1^1) \\
&=&{\frac{\sqrt{2}}{\sqrt{1+r}}}F_{12}(\psi )D_1(\psi )(x_1^1) \\
&=&{\frac{\sqrt{1+r}}{\sqrt{2}}}F_{12}(\psi )(x_1^1) \\
&=&{\frac{1+r}2}x_2^1
\end{eqnarray}
Similarly, one can show that 
\begin{equation}
\Vert x_2^2\Vert ^2=1
\end{equation}
and 
\begin{equation}
D_2(\psi )(x_2^2)={\frac{1-r}2}x_2^2
\end{equation}
Hence this shows that $x_2^1$, respectively $x_2^2,$ is a normalized
eigenvector of $D_2(\psi )$ with eigenvalue ${\frac{1+r}2},$ respectively ${%
\frac{1-r}2}$. From this follows that $D_2(\psi )$ has the form 
\begin{equation}
D_2(\psi )={\frac 12}\left( 
\begin{array}{cc}
1+r & 0 \\ 
0 & 1-r
\end{array}
\right)
\end{equation}
in the base $x_2^1,x_2^2$. Let us write now $\psi $ in the base $%
\{x_1^1\otimes x_2^1,x_1^1\otimes x_2^2,x_1^2\otimes x_2^1,x_1^2\otimes
x_2^2\}$ of ${\mathbb C}_1^2\otimes {\mathbb C}_2^2$, hence 
\begin{equation}
\psi =ax_1^1\otimes x_2^1+bx_1^1\otimes x_2^2+cx_1^2\otimes
x_2^1+dx_1^2\otimes x_2^2
\end{equation}
We have then 
\begin{eqnarray}
F_{12}(\psi )(x_1^1) &=&ax_2^1+bx_2^2={\frac{\sqrt{1+r}}{\sqrt{2}}}x_2^1 \\
F_{12}(\psi )(x_1^2) &=&cx_2^1+dx_2^2={\frac{\sqrt{1-r}}{\sqrt{2}}}x_2^2
\end{eqnarray}
which shows that 
\begin{eqnarray}
a &=&{\frac{\sqrt{1+r}}{\sqrt{2}}} \\
b &=&0 \\
c &=&0 \\
d &=&{\frac{\sqrt{1-r}}{\sqrt{2}}}
\end{eqnarray}
and hence 
\begin{equation}
\psi ={\frac{\sqrt{1+r}}{\sqrt{2}}}x_1^1\otimes x_2^1+{\frac{\sqrt{1-r}}{%
\sqrt{2}}}x_1^2\otimes x_2^2
\end{equation}
which is the Schmidt diagonal form of $\psi $ adapted to our sphere model of
the spin ${\frac 12}$.

\subsection{Non collapse measurement}

Let us now see what the result of a non collapse measurement is on the
density state, using Luder's formula. Let us write the state in the Schmidt
diagonalization form: 
\begin{equation}
\left| \psi \right\rangle ={\frac{\sqrt{1+r}}{\sqrt{2}}}x_1^1\otimes x_2^1+{%
\frac{\sqrt{1-r}}{\sqrt{2}}}x_1^2\otimes x_2^2
\end{equation}
and choose coordinates such that $x_1^1=\left( 
\begin{array}{l}
1 \\ 
0
\end{array}
\right) ,x_1^2=\left( 
\begin{array}{l}
0 \\ 
1
\end{array}
\right) $ in ${\mathbb C_1^2}$ and $x_2^1=\left( 
\begin{array}{l}
1 \\ 
0
\end{array}
\right) ,x_2^2=\left( 
\begin{array}{l}
0 \\ 
1
\end{array}
\right) $ in $\Bbb{C}_2^2.$ The density state $\rho \left( \psi \right) $
corresponding with the pure state $\left| \psi \right\rangle $ is given by: 
\begin{eqnarray}
\rho \left( \psi \right)  &=&\left| \psi \right\rangle \left\langle \psi
\right|  \\
&=&\frac{1+r}2\left( 
\begin{array}{ll}
1 & 0 \\ 
0 & 0
\end{array}
\right) \otimes \left( 
\begin{array}{ll}
1 & 0 \\ 
0 & 0
\end{array}
\right) +\frac{\sqrt{1-r^2}}2\left( 
\begin{array}{ll}
0 & 1 \\ 
0 & 0
\end{array}
\right) \otimes \left( 
\begin{array}{ll}
0 & 1 \\ 
0 & 0
\end{array}
\right)  \\
&&+\frac{\sqrt{1-r^2}}2\left( 
\begin{array}{ll}
0 & 0 \\ 
1 & 0
\end{array}
\right) \otimes \left( 
\begin{array}{ll}
0 & 0 \\ 
1 & 0
\end{array}
\right) +\frac{1-r}2\left( 
\begin{array}{ll}
0 & 0 \\ 
0 & 1
\end{array}
\right) \otimes \left( 
\begin{array}{ll}
0 & 0 \\ 
0 & 1
\end{array}
\right) 
\end{eqnarray}
The projector operator for a measurement along direction $\left( \theta
,\phi \right) $ is given by $P(\theta ,\phi )=D(1,\theta ,\phi ),$ i.e., 
\begin{equation}
P(\theta ,\phi )=D(1,\theta ,\phi )={\frac 12}\left( 
\begin{array}{cc}
1+\cos \theta  & \sin \theta e^{-i\phi } \\ 
\sin \theta e^{i\phi } & 1-\cos \theta 
\end{array}
\right) 
\end{equation}
and its orthogonal by $1-P(\theta ,\phi ),$ i.e., 
\begin{equation}
1-P(\theta ,\phi )=D(1,\pi -\theta ,\phi +\pi )={\frac 12}\left( 
\begin{array}{cc}
1-\cos \theta  & -\sin \theta e^{-i\phi } \\ 
-\sin \theta e^{i\phi } & 1+\cos \theta 
\end{array}
\right) 
\end{equation}
To obtain the density state $\rho ^{\prime }\left( \psi \right) $ after a
non collapse measurement we use Luder's formula, with the following result: 
\begin{eqnarray}
\rho ^{\prime }\left( \psi \right)  &=&\left( P(\theta ,\phi )\otimes 
\mathbf{1}\right) \rho \left( \psi \right) \left( P(\theta ,\phi )\otimes 
\mathbf{1}\right) +\left( \left( \mathbf{1}-P(\theta ,\phi )\right) \otimes 
\mathbf{1}\right) \rho \left( \psi \right) \left( \left( \mathbf{1}-P(\theta
,\phi )\right) \otimes \mathbf{1}\right)  \\
&=&\frac{1+r}4\left( 
\begin{array}{ll}
1+\cos ^2\theta  & \cos \theta \sin \theta e^{-i\phi } \\ 
\cos \theta \sin \theta e^{i\phi } & 1-\cos ^2\theta 
\end{array}
\right) \otimes \left( 
\begin{array}{ll}
1 & 0 \\ 
0 & 0
\end{array}
\right)  \\
&&+\frac{\sqrt{1-r^2}}4e^{i\phi }\left( 
\begin{array}{ll}
\cos \theta \sin \theta  & \sin ^2\theta e^{-i\phi } \\ 
\sin ^2\theta e^{i\phi } & -\cos \theta \sin \theta 
\end{array}
\right) \otimes \left( 
\begin{array}{ll}
0 & 1 \\ 
0 & 0
\end{array}
\right)  \\
&&+\frac{\sqrt{1-r^2}}4e^{-i\phi }\left( 
\begin{array}{ll}
\cos \theta \sin \theta  & \sin ^2\theta e^{-i\phi } \\ 
\sin ^2\theta e^{i\phi } & -\cos \theta \sin \theta 
\end{array}
\right) \otimes \left( 
\begin{array}{ll}
0 & 0 \\ 
1 & 0
\end{array}
\right)  \\
&&+\frac{1-r}4\left( 
\begin{array}{ll}
1-\cos ^2\theta  & -\cos \theta \sin \theta e^{-i\phi } \\ 
-\cos \theta \sin \theta e^{i\phi } & 1+\cos ^2\theta 
\end{array}
\right) \otimes \left( 
\begin{array}{ll}
0 & 0 \\ 
0 & 1
\end{array}
\right) 
\end{eqnarray}

\noindent From this, we can calculate $D_1(\psi ),$ i.e., the partial trace
density matrix to ${\mathbb C}_1^2$ and we obtain 
\begin{eqnarray}
D_1(\psi ) &=&\frac{1+r}4\left( 
\begin{array}{ll}
1+\cos ^2\theta  & \cos \theta \sin \theta e^{-i\phi } \\ 
\cos \theta \sin \theta e^{i\phi } & 1-\cos ^2\theta 
\end{array}
\right)  \\
&&+\frac{1-r}4\left( 
\begin{array}{ll}
1-\cos ^2\theta  & -\cos \theta \sin \theta e^{-i\phi } \\ 
-\cos \theta \sin \theta e^{i\phi } & 1+\cos ^2\theta 
\end{array}
\right)  \\
&=&{\frac 12}\left( 
\begin{array}{cc}
1+r\cos ^2\theta  & r\sin \theta \cos \theta e^{-i\phi } \\ 
r\sin \theta \cos \theta e^{i\phi } & 1-r\cos ^2\theta 
\end{array}
\right) 
\end{eqnarray}
This is the same density matrix as we found for a measurement on a single
spin $\frac 12.$

\noindent Also, we can calculate $D_2(\psi ),$ i.e., the partial trace
density matrix to ${\mathbb C}_2^2$ and we find: 
\begin{equation}
D_2(\psi )=\frac 12\left( 
\begin{array}{ll}
1+r & 0 \\ 
0 & 1-r
\end{array}
\right) 
\end{equation}
which is independent of $\left( \theta ,\phi \right) .$ This means that a
noncollapse measurement on one spin does not provoke any change in the
partial trace density matrix of the other spin: the spins behave as
separated entities for noncollapse measurements.

\subsection{Collapse measurement}

Let us now study the effect of a non collapse measurement using the
constraint functions. Again, the state $\psi $ is written in the Schmidt
diagonalization form: 
\begin{equation}
\left| \psi \right\rangle ={\frac{\sqrt{1+r}}{\sqrt{2}}}x_1^1\otimes x_2^1+{%
\frac{\sqrt{1-r}}{\sqrt{2}}}x_1^2\otimes x_2^2
\end{equation}
with $r\in [0,1],$ and we use $\left\{ x_1^1,x_1^2\right\} ,$ respectively $%
\left\{ x_2^1,x_2^2\right\} $ , as basis for $\Bbb{C}_1^2,$ respectively $%
\Bbb{C}_2^2$. These two orthonormal basis are related by the following
expressions:
\begin{eqnarray}
x_2^1 &=&{\frac{\sqrt{2}}{\sqrt{1+r}}}F_{12}(\psi )(x_1^1)  \label{x21} \\
x_2^2 &=&{\frac{\sqrt{2}}{\sqrt{1-r}}}F_{12}(\psi )(x_1^2)  \label{x22}
\end{eqnarray}
which in the sphere representation means that the north pole of the first
sphere is mapped onto the north pole of the second sphere, and the south
pole of the first sphere is mapped to the south pole of the second sphere
(in the bases $\left\{ x_1^1,x_1^2\right\} $ and $\left\{
x_2^1,x_2^2\right\} $ ).

Let us now study the mapping $F_{12}(\psi )$ for the other states. From (\ref
{x21}) and (\ref{x22}) it follows immediately that $F_{12}(\psi )$ does not
preserve the norm. Let us calculate the norm of $F_{12}(\psi )(z)$ for an
arbitrary vector $z=\psi (\theta ,\phi )$: 
\begin{eqnarray}
\Vert F_{12}(\psi )(z)\Vert ^2 &=&\langle F_{12}(\psi )(z),F_{12}(\psi
)(z)\rangle =\frac{1+r}2\cos ^2{\frac \theta 2}+\frac{1-r}2\sin ^2{\frac
\theta 2} \\
&=&\frac 12\left( 1+r\cos \theta \right) 
\end{eqnarray}
If we consider for a moment the angle $\theta $ as a variable, we see that
the square of the norm varies between $\frac{1+r}2$ and $\frac{1-r}2$,
depending on the value of $\theta $. For $\theta =0$, and hence the state
represented by the north pole of the sphere, we have: 
\begin{equation}
\Vert F_{12}(\psi )(z)\Vert ^2=\frac{1+r}2
\end{equation}
and for $\theta =\pi $, and hence the state represented by the south pole of
the sphere, we have: 
\begin{equation}
\Vert F_{12}(\psi )(z)\Vert ^2=\frac{1-r}2
\end{equation}

Not only the norm, but also orthogonality is in general not conserved by $%
F_{12}(\psi )$. Let us consider for example an orthonormal base $\{\psi
_u=\psi (\theta ,\phi ),\psi _{-u}=\psi (\pi -\theta ,\phi +\pi )\}$. We can
use the conjugate linearity of $F_{12}(\psi )$ to obtain:

\begin{eqnarray}
F_{12}(\psi )(\psi _u) &=&\sqrt{\frac{1+r}2}\cos {\frac \theta 2}\ e^{i{%
\frac \phi 2}}\ x_2^1+\sqrt{\frac{1-r}2}\sin {\frac \theta 2}\ e^{-i{\frac
\phi 2}}\ x_2^2 \\
F_{12}(\psi )(\psi _{-u}) &=&\sqrt{\frac{1+r}2}i\sin {\frac \theta 2}\ e^{i{%
\frac \phi 2}}\ x_2^1-i\sqrt{\frac{1-r}2}\cos {\frac \theta 2}\ e^{-i{\frac
\phi 2}}\ x_2^2
\end{eqnarray}
Therefore, for $0\neq \theta \neq \pi $ we find that orthogonal states are
mapped onto orthogonal states iff:

\begin{eqnarray}
\langle F_{12}(\psi )(\psi _u),F_{12}(\psi )(\psi _{-u})\rangle  &=&\frac{1+r%
}2i\cos {\frac \theta 2}\sin {\frac \theta 2}-\frac{1-r}2i\cos {\frac \theta
2}\sin {\frac \theta 2}=0 \\
&\Leftrightarrow &\frac{1+r}2=\frac{1-r}2 \\
&\Leftrightarrow &r=0
\end{eqnarray}
Translated on the sphere this gives that diametrical opposite points are
mapped to diametrical opposite points only in the special case $r=0,$
(except the north and south pole which are always mapped onto the north and
south pole of the second sphere).

We consider now the following situation. Take vector $\psi _{v_1}=\psi
(\theta _{v_1},\phi _{v_1})$ representing the point $v_1(\theta _{v_1},\phi
_{v_1})$ on the sphere. Consider now: 
\begin{equation}
\psi _{v_2}={\frac 1{\Vert F_{12}(\psi )(\psi _{v_1})\Vert }}F_{12}(\psi
)(\psi _{v_1})
\end{equation}
which is the normalized vector. This means that there are $\theta _{v_2}$
and $\phi _{v_2}$ such that: 
\begin{equation}
\psi _{v_2}=\psi (\theta _{v_2},\phi _{v_2})
\end{equation}
We want to find out where the corresponding point $v_2(\theta _{v_2},\phi
_{v_2})$ lies on the sphere. Therefore we compare the inproduct of $\psi
_{v_2}$ with $x_2^1$ with the inproduct of $\psi _{v_1}$ with $x_1^1$. We
have: 
\begin{eqnarray}
\langle \psi _{v_2},x_2^1\rangle  &=&{\frac 1{\Vert F_{12}(\psi )(\psi
_{v_1})\Vert }}\cdot \frac{\sqrt{2}}{\sqrt{1+r}}\cdot \langle F_{12}(\psi
)(\psi _{v_1}),F_{12}(\psi )(x_1^1)\rangle  \\
&=&{\frac 1{\Vert F_{12}(\psi )(\psi _{v_1})\Vert }}\cdot \frac{\sqrt{2}}{%
\sqrt{1+r}}\cdot \langle \psi _{v_1},D_1(x_1^1)\rangle ^{*} \\
&=&{\frac 1{\Vert F_{12}(\psi )(\psi _{v_1})\Vert }}\cdot \frac{\sqrt{1+r}}{%
\sqrt{2}}\langle \psi _{v_1},x_1^1\rangle ^{*} \\
&=&\sqrt{\frac{1+r}{1+r\cos \theta _{v_1}}}\cdot \langle \psi
_{v_1},x_1^1\rangle ^{*}
\end{eqnarray}
Only in the case when $r=0$ (i.e., the singlet state) we have that the
inproducts are equal (and consequently, antipodal points on the sphere are
mapped to antipodal points, as mentioned before).

An interesting case is when $\theta _{v_1}={\frac \pi 2}$. Then we find: 
\begin{equation}
\langle \psi _{v_2},x_2^1\rangle =\sqrt{1+r}\cdot \langle \psi
_{v_1},x_1^1\rangle ^{*}
\end{equation}
and 
\begin{equation}
\langle \psi _{v_1},x_1^1\rangle ^{*}={\frac 1{\sqrt{2}}}e^{-i{\frac{\phi
_{v_1}}2}}
\end{equation}
To see what this gives on the sphere, we use the following formula: 
\begin{equation}
{\frac{{1+u(\theta ^{\prime },\phi ^{\prime })\cdot u(\theta ,\phi )}}2}%
=|\langle \psi (\theta ^{\prime },\phi ^{\prime }),\psi (\theta ,\phi
)\rangle |^2  \label{productrelation}
\end{equation}
for $\psi (\theta ^{\prime },\phi ^{\prime })=\psi _{v_2}$ (hence ${u(\theta
^{\prime },\phi ^{\prime })=v}_2(\theta _{v_2},\phi _{v_2})$ ) and $\psi
(\theta ,\phi )=x_2^1$. So we get: 
\begin{equation}
{\frac{{1+{v}_2(\theta _{v_2},\phi _{v_2})\cdot u(\theta ,\phi )}}2}=\frac{%
1+r}2
\end{equation}
and as a consequence: 
\begin{equation}
{{v}_2(\theta _{v_2},\phi _{v_2})}\cdot u(\theta ,\phi )=r
\end{equation}
This means that on the sphere, the elements of the equator are mapped onto a
cone that makes an angle $\beta $ with the north south axis of the second
sphere, such that: 
\begin{equation}
\cos \beta =r
\end{equation}
And indeed, only for $r=0$ this is again an equator (and hence conserving
the angle between the elements of the equator and the north pole). For $r\in
\left] 0,1\right[ $ we get a cone with an angle $0<\beta <{\frac \pi 2}$,
which means that the equator has `raised' to the north. For $r$ approaching $%
1$ the sphere is stretched more and more to the north pole of the second
sphere. Remember that in this limit case the superposition state becomes a
product state, and this fits with the fact that for product states indeed
the map $F_{12}(\psi )$ ${}$maps the first element of the product to the
second.

To see the general scheme we use

\begin{equation}
\langle \psi _{v_2},x_2^1\rangle =\sqrt{\frac{1+r}{1+r\cos \theta _{v_1}}}%
\cdot \langle \psi _{v_1},x_1^1\rangle ^{*}
\end{equation}
in the relation (\ref{productrelation}) to obtain: 
\begin{eqnarray}
{\frac{{1+{v}_2(\theta _{v_2},\phi _{v_2})\cdot u(\theta ,\phi )}}2} &=&%
\frac{1+r}{1+r\cos \theta _{v_1}}{\cos ^2{\frac{\theta _{v_1}}2}} \\
&=&\frac{1+r}{1+r\cos \theta _{v_1}}\frac{1+{\cos }\theta _{v_1}}2
\end{eqnarray}
which yields 
\begin{equation}
{{v}_2(\theta _{v_2},\phi _{v_2})}\cdot u(\theta ,\phi )={\frac{r+{\cos }%
\theta _{v_1}}{1+r\cos \theta _{v_1}}}  \label{endresult}
\end{equation}

From formula (\ref{endresult}) it follows that straight lines through the
center of the left sphere are mapped onto straight lines through the point $%
u\left( r,0,0\right) $ along the north south axis in the second sphere,
which gives a nice geometrical representation of this `stretching' on the
second sphere (see Figure 4). This also shows that indeed only for $r=0$
antipodal points of the first sphere are mapped onto antipodal points of the
second sphere.

\begin{figure}[htpb]
\centering
\includegraphics{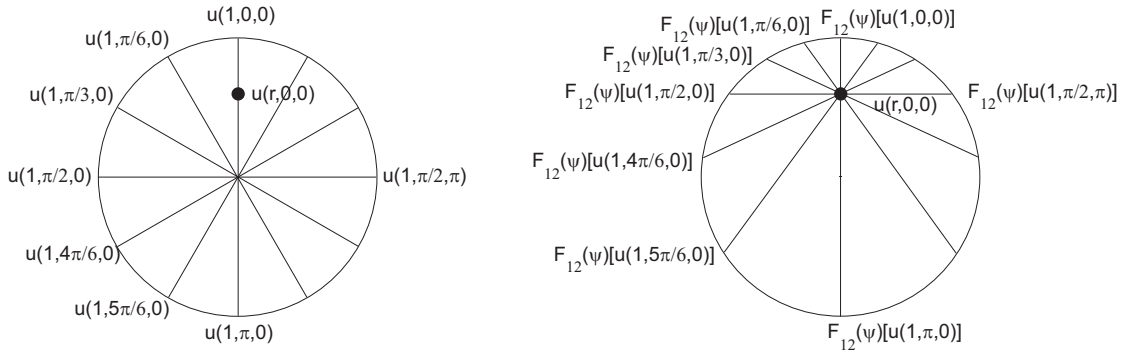}
\caption{Straight lines through the center of the left sphere are mapped onto straight lines through the point $u\left( r,0,0\right) $ along the north south axis in the second sphere.}
\end{figure}

\section{Conclusions}

We have studied the quantum entity consisting of two entangled spin $\frac 12
$ which in standard quantum mechanics is described in the tensor product of
the two dimensional complex Hilbert spaces that describe the single spins.
We have introduced `constraint functions' that describe the behavior of the
state of one of the spins if measurements are executed on the other spin. By
making use of the sphere model representation for the spin $\frac 12$ 's
that was developed in Brussels, we studied in detail the entanglement for an
arbitrary tensor product state, which is not necessarily the singlet state.

We considered two types of measurements: (1) noncollapse measurements, of
which the action on a mixture of states is described by Luder's formula, and
(2) collapse measurements, of which the action is described by Von Neumann's
formula. Our result is that (1) an arbitrary noncollapse measurement on one
spin does not provoke any change in the partial trace density matrix of the
other spin: the spins behave as separated entities for noncollapse
measurements; (2) an arbitrary collapse measurement on one spin provokes a
rotation and a `stretching' on the other spin, which gives a nice
geometrical representation of how entanglement works as an internal
constraint. We conclude by remarking that our study is a further elaboration
of earlier studies of the entanglement influence as constraint, more
specifically to be found in \cite{aerts1991hpa,coecke1998}.

\section{Acknowledgments}

Ellie D'Hondt is Research Assistant of the Fund for Scientific Research -
Flanders (Belgium)(F.W.O. - Vlaanderen). Bart D'Hooghe is a Postdoctoral
Fellow of the Fund for Scientific Research - Flanders (Belgium)(F.W.O. -
Vlaanderen).

\end{document}